\documentclass[referee,traditabstract]{aa} 
\usepackage{graphicx}
\usepackage{natbib}
\bibpunct{(}{)}{;}{a}{}{,}
\usepackage{txfonts}
\begin{document}
%
%\title{ Detection of the ellipsoidal and the relativistic
 %beaming effect induced by the 
 %massive-planet/brown-dwarf CoRoT-3
%}
\title{Kepler KOI-13.01 --- Detection of beaming and ellipsoidal modulations pointing to a massive hot Jupiter 
}
\author{
        T.\ Mazeh           \inst{1} 
\and G. Nachmani        \inst{1} 
\and G. Sokol               \inst{1} 
\and S.\ Faigler         \inst{1} 
\and S.\ Zucker        \inst{2} 
\\
}
\institute{ 
School of Physics and Astronomy, 
Raymond and Beverly Sackler Faculty of Exact Sciences,\\ 
Tel Aviv University, Tel Aviv  69978, Israel
%\email{mazeh@post.tau.ac.il}
\and
Department of Geophysics and Planetary Sciences, 
Raymond and Beverly Sackler Faculty of Exact Sciences,
Tel Aviv University, Tel Aviv  69978, Israel
}
\date{Received / Accepted }

\abstract{ 
KOI-13 was presented by the Kepler team as a candidate for having a giant planet --- KOI-13.01, with orbital period of $1.7$ d and transit depth of $\sim$$0.8$\%.
We have analyzed the Kepler Q2 data of KOI-13, 
%{\bf 
which was publicly available at the time of the submission of this paper, and derived the amplitudes of the beaming, ellipsoidal and reflection modulations --- $8.6\pm 1.1$, $66.8\pm 1.6$ and $72.0\pm1.5$ ppm (parts per million), respectively. 
After the paper was submitted, Q3 data were released, so we repeated the analysis with the newly available light curve. The results of the two quarters were quite similar.
%}
From the amplitude of the beaming modulation we derived a mass of $10\pm2 \, M_{Jup}$ for the secondary, suggesting that KOI-13.01 was a massive planet, with one of the largest known radii. 
We also found in the data a periodicity of unknown origin with a period of $1.0595$ d and a peak-to-peak modulation of $\sim$$60$ ppm.  
%{\bf
The light curve of Q3 revealed a few additional small-amplitude periodicities with similar frequencies.
%} 
\\
It seemed as if the secondary occultation of KOI-13 was slightly deeper than the reflection peak-to-peak modulation by $16.8\pm4.5$ ppm. 
If real, this small difference was a measure of the thermal emission from the night side of KOI-13.01.  We estimated the effective temperature to be 
$2600\pm150$ K, using a simplistic black-body emissivity approximation. We then derived the planetary geometrical and Bond albedos as a function of the day-side temperature. Our analysis suggested that the Bond albedo of KOI-13.01 might be substantially larger than the geometrical albedo. 
}
\authorrunning{Mazeh et al.}  
\titlerunning{KOI-13.01 --- a massive planet}	       
\maketitle

%============================
\section{Introduction}        
\label{introduction}                            
%============================
The Kepler team \citep{borucki11} listed KOI-13 (KIC 9941662)  as a candidate for having a giant planet with an orbital period of $1.7$ d and a transit depth of $\sim$$0.8$\%
(\citet{rowe11} discussed this system in an abstract). 
The planet got a Kepler vetting score of '2', which, according to the classification of Borucki et al., meant that the planet was a "strong probability candidate, cleanly passes tests that were applied". This score was only next to that of the planets confirmed with radial-velocity (RV) follow-up observations, which got a vetting score of '1', and were then christened an official planet sequential Kepler numbering, like Kepler-1a, Kepler-2a, etc \citep{borucki10}.  

The Kepler Input Catalog \citep{brown11} stellar temperature of KOI-13, as appeared in \citet{borucki11}, is 8448 K. Stars with such temperature have very few, usually wide, stellar lines in their spectra and therefore measuring their RV is difficult \citep[e.g.,][]{galland05}. It seemed therefore that the planet around KOI-13  --- KOI-13.01, was doomed to stay with the '2' vetting score, and might not reach the official Kepler-planet exclusive club. 

This is unfortunate, because \citet{borucki11}, based on the KIC stellar mass and radius and the analysis of the Kepler light curve, estimated the planet candidate radius to be $1.86 \pm 0.003 R_{Jup}$. This is one of the largest derived radii of transiting planets. According to the Extrasolar Planet Encyclopedia (http://exoplanet.eu/), as of July 2011 only WASP-17b \citep{anderson11} and HAT-P-32b \citep{hartman11} have larger radii, 
of about 
$2\,R_{Jup}$. These two confirmed planets have masses of $0.5$ and $0.9\,M_{Jup}$, respectively, and therefore are some of the most bloated, low-density planets known.  It would be interesting to have an estimate of the mass of KOI-13.01 and see if it is also one of those bloated planets.

In fact, very recently \citet{szabo11} used the derived large radius to argue that KOI-13.01 was not a planet at all. Szab{\'o} et al. noticed that KOI-13 was the known BD+46\,2629 visual double star, and convincingly proved, by analyzing the publicly available Kepler images, and by obtaining ground-based lucky imaging photometry, that the planet candidate orbited the brighter of the two stars --- KOI-13A. They also showed that the light coming from KOI-13B contributed 45\% of the light of the system in the Kepler band. Furthermore, Szab{\'o} et al.\ obtained a high-resolution spectrum of KOI-13 and derived its mass, radius, temperature and stellar rotation --- $v\sin i_{rot}$, where $i_{rot}$ was the inclination angle of the stellar rotation axis. The latter was estimated to be $65\pm10$ km/s. They also discovered that the transit had an asymmetric part, on the order of 100 ppm (parts per million), and suggested that this was caused by some obliquity of the orbit and high gravity darkening of the stellar surface, due to its high temperature and fast rotation.  After correcting for the dilution factor and the gravity darkening effect, \citet{szabo11} derived the radius of KOI-13.01 to be $2.2\pm0.1\,R_{Jup}$. They used this value to argue that KOI-13.01 was a brown dwarf, as the theoretical models of planet evolution they cited \citep{fortney07} could not account for such a large radius.

The conclusion of Szab{\'o} et al. seems somewhat premature, given the fact that we know of HAT-P-33b and WASP-17b, two {\it bona fide} planets with RV mass determination that have radii $\sim$$2\,R_{Jup}$. We would guess that the theoretical border of the allowed range of planetary radii is not so sharp. It is therefore even more desirable to have some observational estimate of KOI-13.01 mass, so we can confront the theory with the observations.      

One way to estimate the mass of KOI-13.01 is to use the Doppler beaming (sometimes called Doppler boosting) \citep{loeb03,zucker07} and the ellipsoidal \citep[see a review by][]{mazeh08} effects. Both effects cause a modulation of the stellar brightness with the orbital period (beaming) and its first overtone (ellipsoidal).
To first approximation, both effects depend linearly on the secondary mass and therefore detecting these two effects and measuring their amplitudes can, in principle, yield a mass estimate for the planet candidate. Such an approach was applied recently by \citet{carter11} to KIC 10657664 --- an A-star with a white-dwarf secondary, although the results from the two modulations were substantially different. We suspect that the difference came from the fact that the first-order approximation for the ellipsoidal effect is not accurate enough. 
   
The beaming effect increases (decreases) the brightness of any light source approaching (receding from) the observer, enabling the derivation of the stellar Doppler radial-velocity modulation from its precise photometry \citep{loeb03,zucker07}. The amplitude of the Doppler beaming is on the order of  $4V_{rel}/c$, where $V_{rel}$ is the radial velocity of the source relative to the observer and $c$ is the speed of light \citep[e.g.,][]{rybicki79}. The ellipsoidal  effect \citep[e.g.,][]{morris85}, due to
the tidal distortion of the star by the gravity of its companion, is a well known effect in short-period binaries. 

Given the stellar mass and radius and the orbital period of KOI-13, the expected amplitudes of the beaming and the ellipsoidal effects \citep[e.g.,][]{faigler11a} are $\sim$$1.5(m_p/M_{Jup})$ and $\sim$$15(m_p/M_{Jup})$ ppm. If indeed the secondary of KOI-13 is a brown dwarf, then the beaming amplitude is expected to be on the order of 30 ppm or even larger and the ellipsoidal effect should be ten times larger. Such periodic modulations, although much smaller than in the aforementioned case of KIC 10657664 , can be detected in the Kepler data, the ellipsoidal effect in particular. 

The two effects were already detected in a similar case --- CoRoT-3, where a brown dwarf with a mass of $22\,M_{Jup}$ was discovered to orbit an F-type star with a period of 4.3 days \citep{deleuil08}. In that case \citet{mazeh10} detected the beaming and the ellipsoidal effects with amplitudes of  $27\pm9$ and 
$59\pm9$ ppm, respectively, consistent with the mass previously known from the RV measurements. There is no reason why the Kepler publicly available data, which has a similar time span to that of CoRoT-3, would not reveal periodic modulations with similar amplitudes.  In fact, \citet{welsh10} identified in the Kepler data the ellipsoidal effect induced by the known planet HAT-P-7b.

\citet{szabo11} presented in their analysis of KOI-13 another smooth periodic modulation, well observed in short-period binaries --- the reflection/emission modulation
(referred to hereafter as the reflection modulation), induced by the stellar light reflected by the day side of the secondary together with its thermal emission
\citep[e.g.,][]{cowan11}. In the folded light curve presented by \citet{szabo11} the occultation of the secondary by the primary star was also a very prominent feature. Given the secondary radius, the reflection phase modulation and the occultation depth can be used to put some constrains on the secondary albedo and temperature \citep[e.g.,][]{snellen09,demory11}. The derivation of these parameters is another opportunity to explore the nature of KOI-13.01.

In this paper we present our analysis of the Q2 publicly available Kepler data of KOI-13, detecting and deriving the amplitudes of the beaming, ellipsoidal and reflection modulations (Section~2).  
%{\bf 
After this paper was submitted, Q3 data were released, so we repeated the analysis with the Q3 data, the results of which are presented in Section~3. The results of the two quarters are quite similar.
%} 
We use the amplitude of the beaming modulation to estimate the mass of KOI-13.01 and show that it is probably a massive planet (Section~4). We use the derived amplitude of the reflection and the occultation depth to discuss the albedo and temperature of KOI-13.01 (Section~5). Section~6 summarizes our findings. 

%============================
\section{Data analysis --- Q2 data}        
\label{analysis}                            
%============================

The analysis we present in this section is based on the long cadence, raw Q2 Kepler data, spanning 88.9 days, from June 2009 till September 2009. These data were publicly available when we submitted the paper. We decided not to use the Q0 and Q1 data, as the transits of KOI-13 in that part of the light curve seemed inconsistent with the rest of the publicly available data. 

\subsection{Preprocessing the light curve}
%------------------------------------------

We 'cleaned' the data in the same way \citet{mazeh10} preprocessed the CoRoT-3 light curve. This includes: 

\begin{itemize}

\item
Transits removal.  Measurements taken during and around the 50 transits 
of KOI-13 were removed from the analysis (but see below the analysis of the transits themselves.)

\item
Dilution correction. We subtracted from all data points $0.45$ times their median, which is the dilution factor according to \citet{szabo11}.
 
\item
Outlier removal. We identified 58 outliers by 
calculating the running median and r.m.s.\ of 31 measurements around each point, and rejecting
measurements that were $3\sigma$ ($4\sigma$) or more higher (lower) than their
corresponding median. This left a total of 3555 data points.

\item

Long-term detrending with a cosine filter.
We used a discrete cosine transform \citep{ahmed74},  
adapted to the unevenly spaced data we had in hand (for details see \citet{mazeh10}).
We fitted the data with a linear combination of the first $51$ low-frequency cosine
functions, the last of which is with a period of $4.351$ d, and subtracted those functions from the data.

\end{itemize}

%{\bf 
We took special measures to ensure that the results did not depend on the long-term detrending. First, we decreased the number of low-frequency cosine functions till all the derived amplitudes converged. This left us with only 51 frequencies removed by the detrending. Second, we tried an independent smoothing algorithm, with a time span of 3 orbital periods, which yielded almost identical amplitudes. We therefore concluded that our detrending is a robust step of the analysis.
%}   

Figure~1 shows the light curve before and after the removal
of the long-term trend. 
The r.m.s.\ of the cleaned light curve
is 80 ppm. 

%---------------------------
% Figure 1 %
%---------------------------
\begin{figure*}
\centering
\resizebox{15cm}{10cm}
{\includegraphics{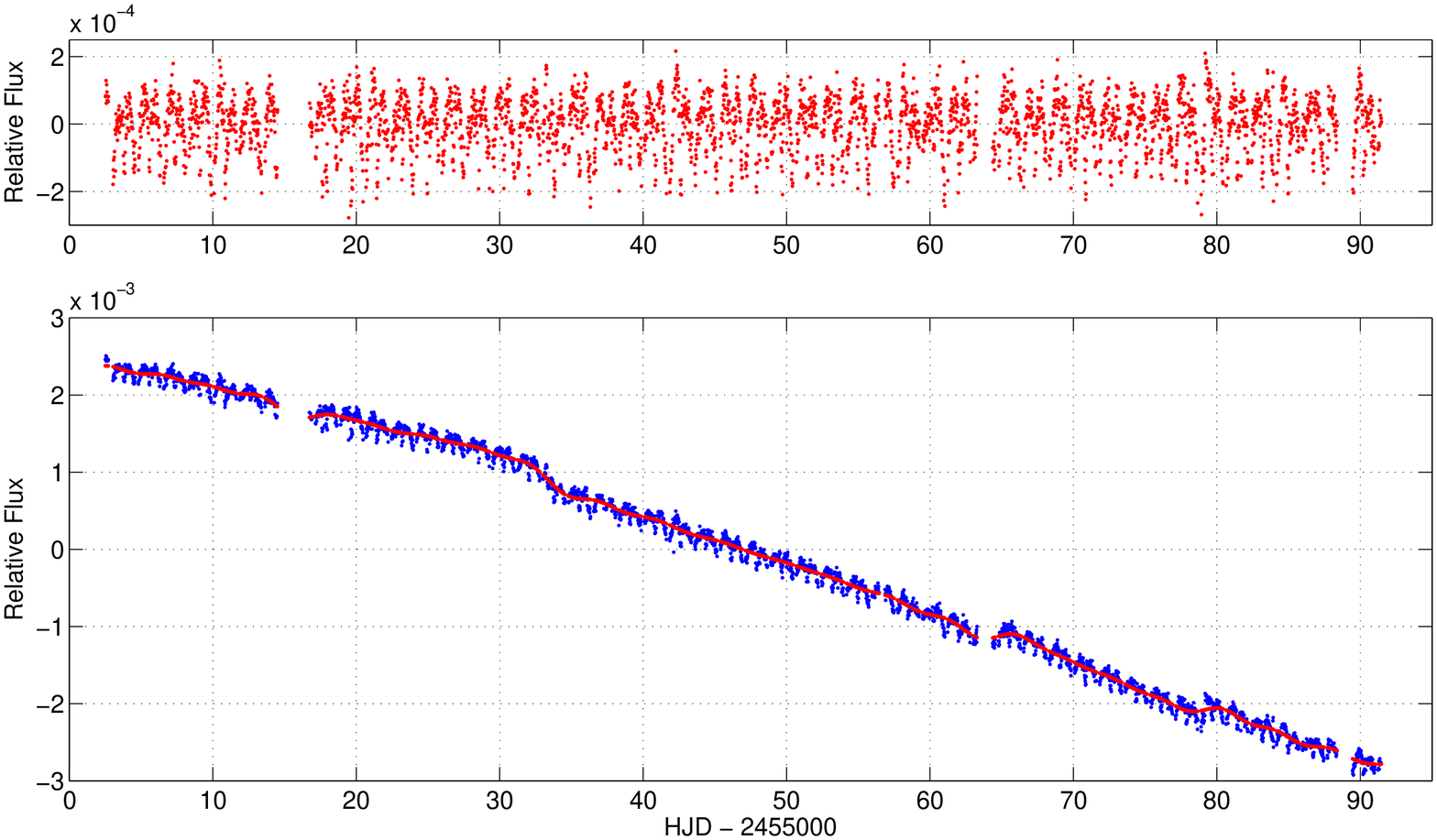}}
\caption{The Q2 raw light curve of KOI-13, before (lower panel) and after (upper panel) detrending. The plotted flux is relative to the mean flux, corrected for the contribution of KOI-13B.  The long-term model is presented by the red line. Note the different scales of the two panels. 
}
\label{detrend}
\end{figure*}
%-------------------------------------------------------------

\subsection{Fitting amplitudes for the beaming, ellipsoidal,  and
reflection effects}
%---------------------------------------------------------------------------------
We fitted the cleaned data with a model that included the ellipsoidal,
beaming and reflection effects \citep[hereafter the BEER model, following][]{faigler11a}. We approximated the beaming and the ellipsoidal modulations by pure sine/cosine functions, using mid-transit, $t_{tran}= 2454964.93$ HJD, of \citet{borucki11} 
as phase zero. The beaming effect was presented by a
sine function with the orbital period, and
the ellipsoidal effect by a cosine function with half the orbital
period. The reflection was approximated by the Lambert law \citep[e.g.,][]{demory11}.

In this approximation we expressed the relative stellar
flux modulation
$\Delta F/\bar{F}$, where $\bar{F}$ was the {\it averaged} detrended, corrected for dilution, stellar flux,  
as a function of the orbital phase 
\begin{equation}
\phi=\frac{2\pi}{P_{\rm orb}}(t-t_{tran})\ ,
\end{equation}
where $P_{\rm orb}$ was the orbital period. We then wrote

%beaming  
\begin{equation} 
\frac{\Delta F_{\mathrm{beam}}(t)}            {\bar{F}}  =
A_{\mathrm{beam}}\sin\phi \ ,
\end{equation} 

%ellip
\begin{equation} 
\frac{\Delta F_{\mathrm{ellip}}(t)}   {\bar{F}} =
-A_{\mathrm{ellip}}\cos2\phi \ ,
\end{equation} 

%refl  
\begin{equation} 
\frac{\Delta F_{\mathrm{refl}}(t)}{\bar{F}} =
-A_{\mathrm{refl}}
\left[2\frac{\sin\phi+(\pi-\phi)\cos\phi}{\pi}\right] \ ,
\end{equation} 
where the coefficients, $A_{\mathrm{ellip}}$, $A_{\mathrm{beaming}}$ and
$A_{\mathrm{refl}}$ were
all positive. Note that we defined $A_{\rm refl}$ to represent half the peak-to-peak variation of the reflection effect, to be consistent with the other two amplitudes.

We fitted the cleaned, detrended light curve of KOI-13 with a
5-parameter {\it linear} model,
%
%  Eq. 5
%----------
\begin{equation}
\mathcal{M}_{{\mathrm{BEER}}}(t)=a_0
+A_{\mathrm{beam}}\sin\phi 
-A_{\mathrm{ellip}}\cos2\phi
-A_{\mathrm{refl}}
\left[2\frac{\sin\phi+(\pi-\phi)\cos\phi}{\pi}\right]
+a_{2s}\sin2\phi \ .
\end{equation}
This was similar to what has been done by \citet{mazeh10} and \citet{faigler11a}, except for the expression of the reflection effect. 
It was important to include the first overtone sine function in the model as a check of our approach. We expected this term to be small, as it had no specific role in the model.
The fitting process could find any value, positive or negative for the five parameters.
However, we did expect $A_{\rm beam}$, $A_{\rm ellip}$ and $A_{\rm refl}$ to be positive and $a_{2s}$ to be close to zero. 

%=========================
% Table 1 %
\begin{table}
\caption{
The best-fit coefficients 
of the three effects and the occultation, using the Q2 and the Q3 data
%\protect\footnotemark[1]
}
%--------------------------------------------
\begin{tabular}{lrrcl}
                           & Q2  \ \ \ \  &     Q3 \ \ \ \ \\
\hline
BEER model:\\
$A_{\rm beam}$  &$ 8.6\pm 1.1$    &   $10.4\pm1.0$       &ppm&   BEaming \\
$A_{\rm ellip}$    &$66.8\pm 1.6$   &    $69.1 \pm1.5$       &ppm&    Ellipsoidal \\
$A_{\rm refl}$     &$72.0 \pm 1.5$  &     $72.0\pm1.4$      &ppm&  Reflection \\
$a_{2s}$             &$ -1.0\pm 1.0$  &      $-3.3\pm1.1$     &  ppm& ---  \\
\hline
Primary transit:\\
$a/R_*$               &$3.16\pm0.08$  &  $3.17\pm0.06$     &       & Semi-major axis to stellar radius\\
$r_p/R_*$         &$0.0908\pm0.0005$&$0.0907\pm0.0005$         &         & Planetary to stellar radii\\
$b$                     &$0.75\pm0.01$  & $0.75\pm0.01$          &        & Impact parameter\\
\hline
Secondary occultation:&\\
depth                 &$163.8\pm3.8$&   $157.8\pm3.5$          &      ppm&  \\
$\Delta\phi$  &$-0.0007\pm0.0005$ &   $-0.0012\pm0.0005$        &\\
\hline
\end{tabular}
%\footnotetext[1]{All values are corrected for the 45\% light dilution of KOI-13B}
\label{table_coeff}
\end{table}
%--------------------------------------------------------------
   
\begin{figure*}
\centering
\resizebox{15cm}{10cm}
{\includegraphics{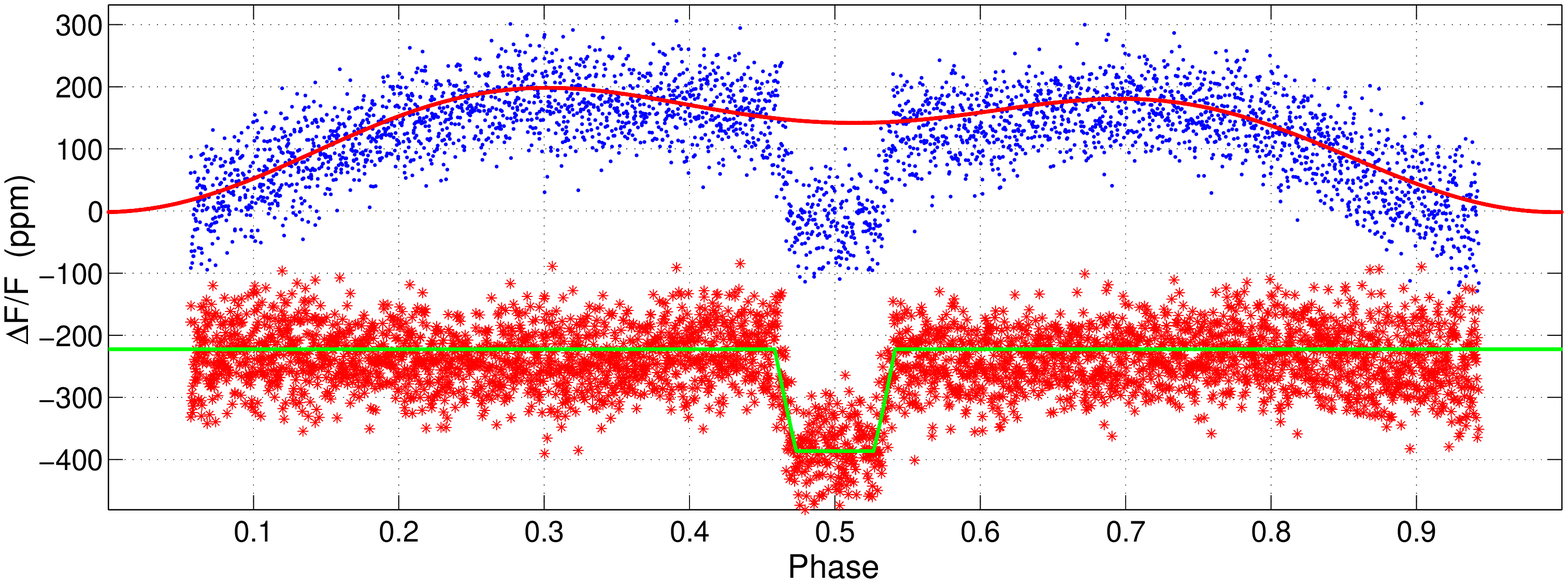}}
\caption{The folded cleaned Q2 light curve of KOI-13 with the best-fit BEER model. The residuals are plotted at the bottom of
the figure, with a trapezoid fitting of the secondary occultation.
}
\label{model_fit}
\end{figure*}

Figure~\ref{model_fit} shows the best-fit model and Table~\ref{table_coeff} lists the resulting amplitudes, after applying the dilution correction due to KOI-13B. The table includes also the results of fitting the model to Q3, as discussed in the next section. All three amplitudes come out to be positive, as expected.  In addition, the fourth coefficient is substantially smaller than the smallest amplitude, and is consistent with zero, as expected. The errors of the four coefficients are around 1--2 ppm, attesting to the high precision of the Kepler satellite.
 
The ellipsoidal modulation, with two peaks, one at a phase of $0.25$ and the other at $0.75$, is very prominent in the figure. So is the reflection effect, which causes the folded light curve to be brighter at phase $0.5$ than at phase $0$. The beaming modulation, which causes the peak at $0.75$ to be lower than the one at $0.25$, is quite small but nevertheless noticeable even by eye, with statistical significance of almost $\sim$$8\sigma$. 

One can also easily notice the secondary occultation at phase $0.5$. We fitted the occultation with 
%{\bf 
a plain trapezoid,
using only two parameters --- the depth and the phase of the occultation. It turned out that the shape of the trapezoid was determined by the three geometrical parameters of the systems --- $a/R_*$, the ratio of the semi-major axis to the stellar radius, $R_*/r_p$ the ratio of the stellar to the planetary radii, and $b$, the impact parameter. Those parameters could have been derived accurately from the planetary {\it transit}. Therefore, although it was not the goal of this study, we nevertheless fitted the long-cadence data of the planetary transit with the \citet{mandel02} model implemented in MATLAB\footnote{Can be found now at www.astro.washington.edu/users/agol/transit.html}, using quadratic limb darkening, with parameters that were found to best fit the data. Our fit used the value of the \citet{mandel02} model at the Kepler timing of each point, and ignored the finite sampling interval of the data.

The best derived values of the three geometrical parameters, used to fit the secondary occultation, are also listed in Table~1. 
%} 
The phase of the occultation is very close to $\phi=0.5$. The difference, denoted in the table as $\Delta\phi$, is consistent with zero, probably indicating a highly circular orbit.   
 
%---------------------------
% Figure 2 %
%---------------------------
%-------------------------------------------------------------

\subsection{Another unexpected periodicity}
%-----------------------------------------------------
In order to look for additional periodicities we calculated the Lomb-Scargle periodogram \citep{scargle82} of the residuals, derived from the preprocessed data by removal of the BEER and the secondary model.  The residuals and their periodogram are plotted in Figure~3.

%---------------------------
% Figure 3 %
%---------------------------
\begin{figure*}
\centering
\resizebox{15cm}{11cm}
{\includegraphics{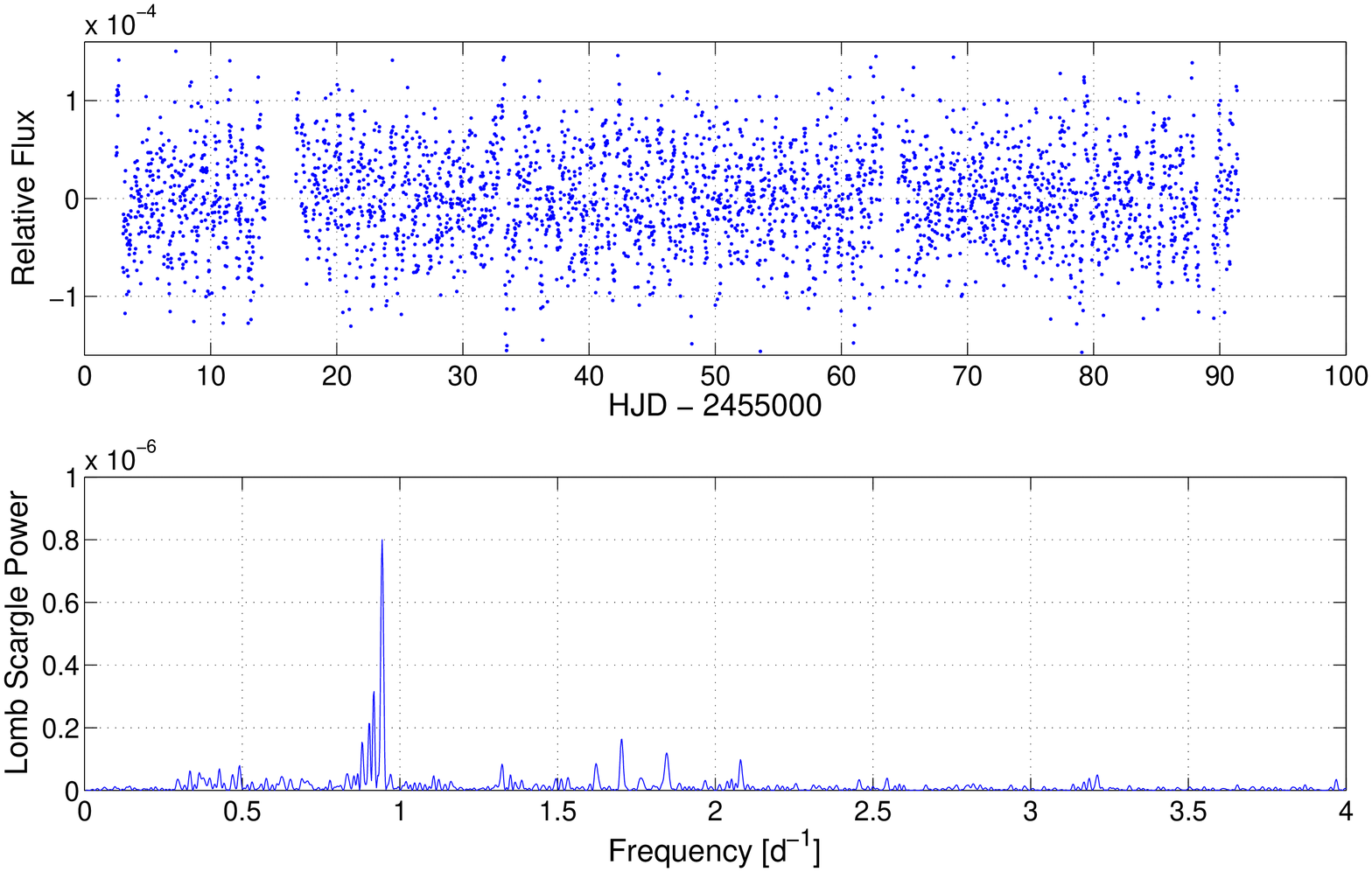}}
\caption{The Q2 residuals of KOI-13, after removing the BEER and the secondary model (upper panel) and their Lomb-Scargle periodogram (lower panel). 
}
\label{LSres}
\end{figure*}

The figure shows a prominent peak at a frequency of $0.9438 \pm 0.001$ d$^{-1}$,  corresponding to a period of $1.0595$ d. Although 
suspiciously close to
the notorious 1-day period, which might point to some telluric
origin, 
we can not see any reason to exclude an astrophysical  
modulation (see further discussion in the next section.) The folded data with the newly found period is shown in Figure~4, which shows that the amplitude of the modulation (peak-to-peak) is $\sim$$60$ ppm.

%---------------------------
% Figure 4 %
%---------------------------
\begin{figure*}
\centering
\resizebox{11cm}{9cm}
{\includegraphics{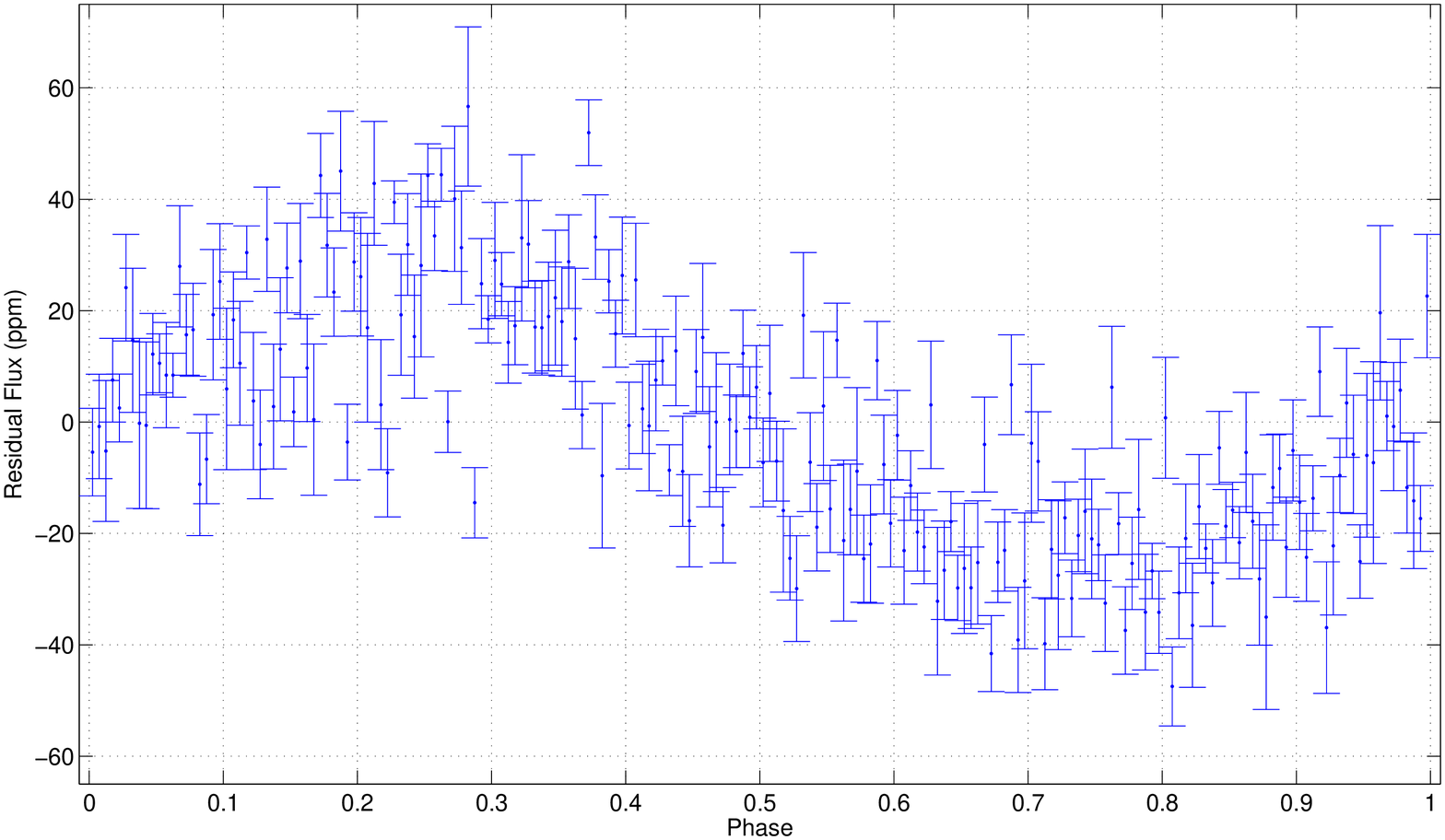}}
\caption{The Q2 residuals of KOI-13, after removing the BEER and the secondary model, folded with the newly found period of $1.0595$ d, and binned into 200
bins. The error bar of each bin presents the scatter of the points in that bin, divided by the square root of the number of points in that bin.
}
\label{mysterious_periodicity}
\end{figure*}

%{\bf
%===============================
%section 3
\section{Analysis of Q3 data}
%===============================

After we submitted this paper, Kepler Q3 data were released, so we repeated the analysis with the newly available raw data. We opted to analyze the two data sets separately, in order to have an independent handle on the errors of the derived parameters. 

The Q3 light curve includes data taken during 89.3 days, from September 2009 to December 2009. The preprocessing analysis was similar to the one of Q2, after which we were left with 3613 data points with an r.m.s. of 77 ppm. We then fitted the BEER and the trapezoid models, the results of which are listed in a separate column of Table~1. 

The newly derived parameters of the three modulations are quite similar to the ones of Q2 data. All differences are within $1.2\sigma$ of their corresponding errors. Even the difference in the secondary occultation depth, $6.0\pm5.3$ ppm, is not significant. Therefore, for the following discussion we adopted the values derived from Q2.

\begin{figure*}
\centering
\resizebox{15cm}{11cm}
{\includegraphics{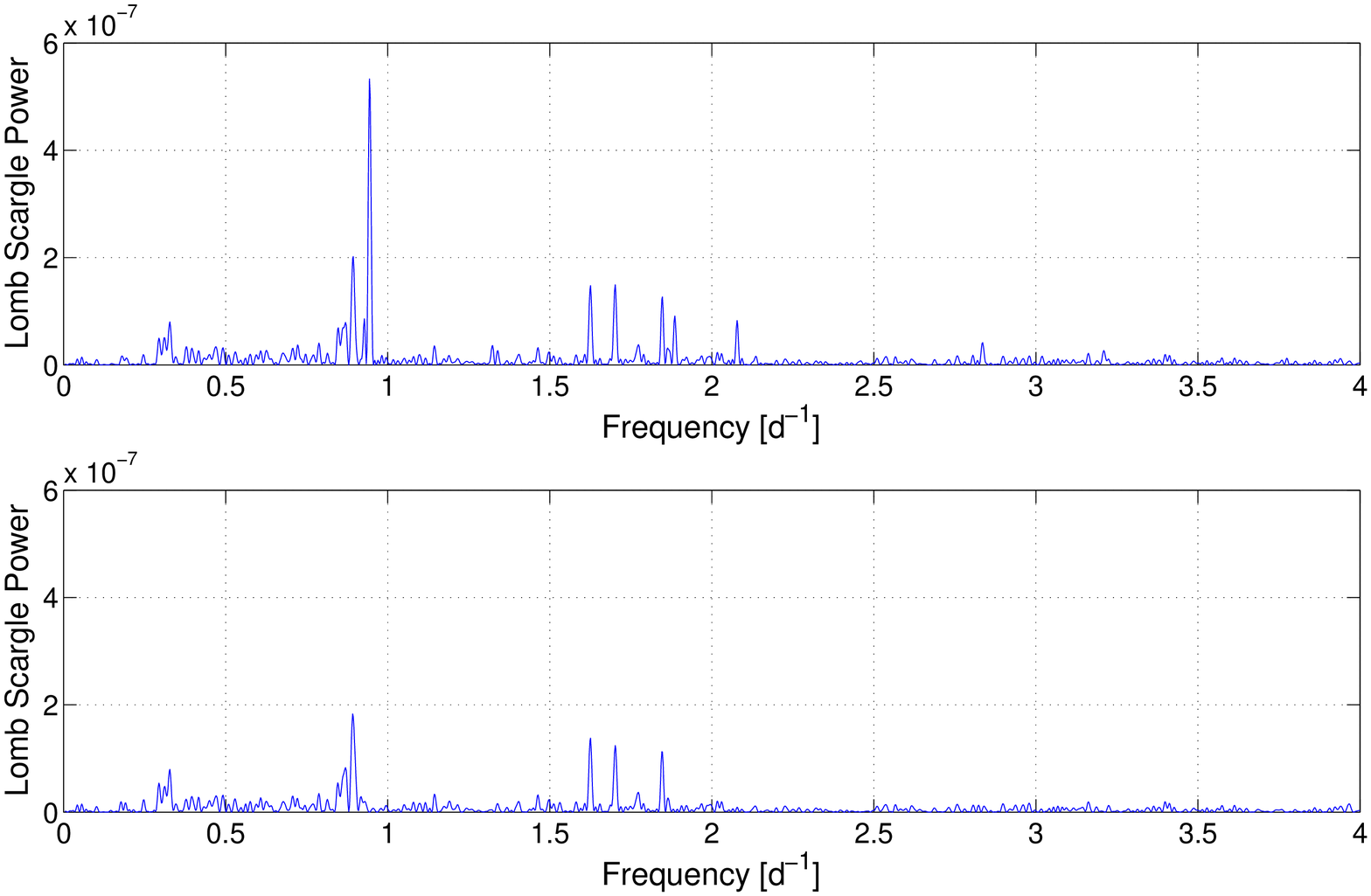}}
\caption{Upper panel: The power spectrum of the {\it Q3} residuals of KOI-13, after removing the BEER and the secondary model. Lower panel: The same, after removing the main periodicity of  $1.0595$ d and its four overtones. For comparison, we kept the same scale for the two panels.
}
\label{Q3power_spectrum}
\end{figure*}

In order to check the additional periodicity of $1.0595$ d found in Q2, we plotted the 
power-spectrum of the residuals  found in the Q3 data.  The derived spectrum shows a few more peaks, as can be seen in the upper panel of Figure~\ref{Q3power_spectrum}. The fact that those additional peaks were not seen in Figure~\ref{mysterious_periodicity} might attest to the better quality of the Q3 dataset. The additional frequencies can be seen better in the lower panel of the figure, which shows the Q3 power spectrum after removing the 1.0595 periodicity with its four overtones.

%---------------------------
% Figure 5 %
%---------------------------
%}

%===============================
%section 4
\section{The mass of KOI-13.01}
%===============================

In this section we derive the mass of the planet candidate KOI-13.01, based on 
\begin{itemize}
\item
the amplitude of the beaming and ellipsoidal effects,
\item
the first order theoretical expectation of the beaming and ellipsoidal effects, as parameterized by \citet{mazeh10} \citep[see also][]{faigler11a}, and
\item
the stellar parameters, as derived by \citet{szabo11}.
\end{itemize}

Using \citet{faigler11a} approximations and notations we can derive the secondary mass from the beaming and ellipsoidal amplitudes: 

%Beaming
\begin{equation}
m_{2, {\rm beam}}
\simeq 
\frac{0.37}{\alpha_\mathrm{beam}} 
\frac{A_\mathrm{beam}}{1\, {\rm ppm}}
\left(\frac{M_*}{M_{\odot}}\right)^{2/3}
\left(\frac{P_\mathrm{orb}}{1 \,{\rm day}}\right) ^{1/3} \, M_{Jup}
\end{equation}
%-----------------

%Ellipsoidal
\begin{equation} 
m_{2, {\rm ellip}}
\simeq
\frac{0.08}{\alpha_\mathrm{ellip}} 
\frac{A_\mathrm{ellip}}{1\, {\rm ppm}}
\left(\frac{R_*}{R_{\odot}}\right)^{-3}
\left(\frac{M_*}{M_{\odot}}\right)^{2}
\left(\frac{P_\mathrm{orb}}{1 \,{\rm day}}\right) ^{2}\, M_{Jup} \ .
\end{equation} 
%------------------ 
%
In the above expressions $M_*$ and $R_*$ are the primary mass and
radius, respectively. The two $m_2$'s present the estimated mass of the secondary from the two amplitudes, and the two $\alpha$'s are numerical factors which are on the order of unity. For completeness, we list the relevant parameters of the system in Table~2. The period value is $1.763582\pm 0.0000014$  d.

%--------------------------------------------------------
\begin{table}
\caption{KOI-13 A parameters, as derived by \citet{szabo11}}
 % \vskip 1pc
\begin{tabular}{lll}
Parameter & Derived value   & Unit  \\
\hline
$M_*$        & $ 2.05 $                   &  $M_\odot$    \\
$R_*$         & $2.55 $                    &  $R_\odot$\\
$T_*$         & $8500\pm400$ \ K    &              \\ 
Dilution factor&1.82\\
\hline
\end{tabular}
\label{table_KOI13A}
\end{table}
%-------------------------------------------------------------

The $\alpha_{beam}$ factor includes one component that originates from the fact that the stellar spectrum is Doppler shifted. To estimate this factor we followed \citet{faigler11b} and numerically shifted spectra from the library of \citet{castelli04} models, which were then integrated over the Kepler bandpass \citep{zucker07}, yielding for each spectrum an $\alpha_{beam}$ factor. The value we adopted for KOI-13,

\begin{equation}
\alpha_\mathrm{beam}=0.63\pm0.06
\end{equation}  
was derived by interpolation between the
available models of the library that were close to the  temperature of KOI-13 given by \citet{szabo11}. We estimated the uncertainty by calculating the
interpolated $\alpha_\mathrm{beam}$ values within the $T_\mathrm{eff}$ error range, and by taking into account the possible gravity and metallicity ranges. All together, the error on  $\alpha_\mathrm{beam}$ amounted to 10\%.

The uncertainty of $\alpha_\mathrm{ellip}$ was much larger. 
We ran numerous models of 
EBOP \citep{etzel80, popper81}, EBAS \citep{tamuz06}, and Wilson-Devinney \citep{wilson71} codes 
to fit the KOI-13 lightcurve and tried to estimate the value of $\alpha_\mathrm{ellip}$ and get a better value than the unity default estimate, to no avail. This might be due to the high gravity darkening and the fast rotation of KOI-13. We therefore adopted a conservative estimate of the numerical factor of 

\begin{equation}
\alpha_\mathrm{ellip}=1^{+1}_{-0.5} \ .
\end{equation}

To complete the error estimate of the secondary mass we needed the uncertainties of the stellar mass and radius of KOI-13. \citet{szabo11} did not give error estimates for their values and therefore we estimated the stellar mass and radius error to be 10\%.

Using the values of Table~2, we got: 

%Beaming
\begin{equation}
m_{2, {\rm beam}}
\simeq 
10\pm2 \, M_{Jup}
%\end{equation}
%-----------------
%
%Ellipsoidal
%\begin{equation}
\ \ \ \ \ \ 
m_{2, {\rm ellip}}
\simeq
4^{+4}_{-2} \, M_{Jup} \ .
\end{equation} 
%------------------ 
%
Due to its large fractional error, the mass derived from the ellipsoidal modulation was not very useful. We therefore adopted the value derived from the beaming modulation. This value  indicated that the secondary mass was very probably in the higher mass range of the planetary regime. 

%===============================
% Section 5
\section{The albedo and temperature of KOI-13.01}
%===============================

The strong insolation of the A-type star that hits KOI-13.01 should heat up its day side to a relatively high temperature. If one assumes that the planet has reached a full synchronization, that all stellar energy is absorbed and no transfer of energy to the night side takes place, then the day-side temperature can be of the order of 3500 K. This is high enough to explain the observed phase modulation with peak-to-peak amplitude of $144\pm1$ ppm. Obviously, according to this approach the night side does not emit any light and therefore the occultation depth should be equal to the phase modulation amplitude. The depth values we derived in the previous sections --- 
$163.8\pm3.8$  for Q2  and $157.8\pm3.5$ for Q3,
indicate that the occultation depth might be slightly deeper. According to the simplistic approach that assumes no light from the night side, the small derived difference between the occultation depth and the phase modulation, denoted here as $\Delta_{NS}$, is necessarily insignificant.   

However, if the observed depth difference is real, we have to assume that it comes from the night side, and therefore it gives us a rare opportunity to explore the night-side thermal emissivity, the Bond and geometrical albedos of the planet, and finally the effective temperatures of the day and the night sides of the planet.

\begin{figure*}
\centering
\resizebox{10cm}{8cm}
{\includegraphics{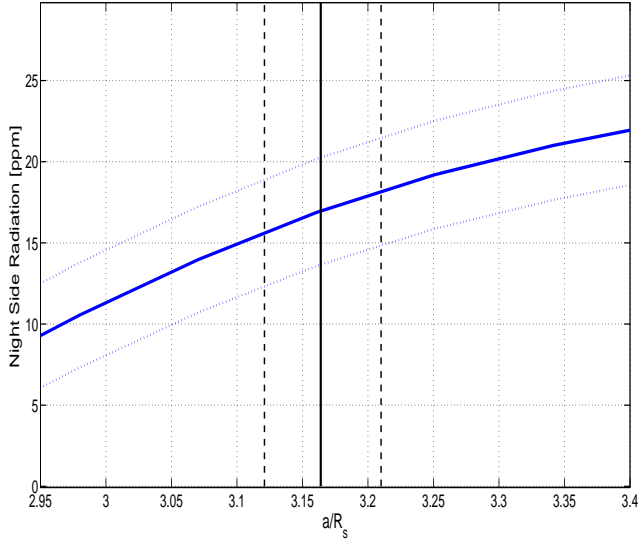}}
\caption{The difference between the occultation depth and the phase modulation, $\Delta_{NS}$, as a function of $a/R_*$, based on Q2 {\it and} Q3 data, combined. The vertical lines present the best-fit $a/R_*$ and its $1\sigma$ error estimation, derived from the planetary transit. 
}
\label{NS}
\end{figure*}

To assess the significance of the non-zero detection of $\Delta_{NS}$ we first note that the figures of Q2 and Q3 in Table~1, taken at face value, yield 
$\Delta_{NS}=16.8\pm 2.8$ ppm.
However, as mentioned in the previous sections, these values depend on the geometrical parameters of the system, derived here by the analysis of the transit. In particular, the trapezoid depth depends on $a/R_*$, the ratio of the semi-major axis to the stellar radius. To further estimate the reality of the $\Delta_{NS}$ detection we plot in Figure~\ref{NS} the weighted average value of $\Delta_{NS}$, obtained from Q2 and Q3, as a function of $a/R_*$. The figure also shows the $1\sigma$ error on $a/R_*$. From the figure we conclude that our best estimate for the night-side emissivity is 

\begin{equation}
\Delta_{NS}=16.8\pm 4.5\ {\rm ppm}\, .
\end{equation}
The figure suggests that the detection of the small difference between the occultation depth and the phase modulation might be real.

%---------------------------
% Figure 6 %
%---------------------------

We now move to exploit the value derived for $\Delta_{NS}$ for better understanding of KOI-13.01 energy budget. To begin, we find that the 
night-side effective temperature, $T_{\rm N}$, needed to account for the observed $\Delta_{NS}$ is $2600\pm150$ K, assuming $r_p/R_*=0.0908\pm 0.0003$ (see Table~1). In the error budget we took into account the uncertainties of the stellar temperature, the stellar and planetary radius ratio, and the error of $\Delta_{NS}$.
The derivation of $T_{\rm N}$ assumes a black-body emissivity of the night side, obviously a very simplistic assumption \citep{cowan11}. A detailed model of the planetary atmosphere \citep[e.g.,][]{fortney08} is well beyond the scope of this paper.

\begin{figure*}
\centering
\resizebox{10cm}{8cm}
{\includegraphics{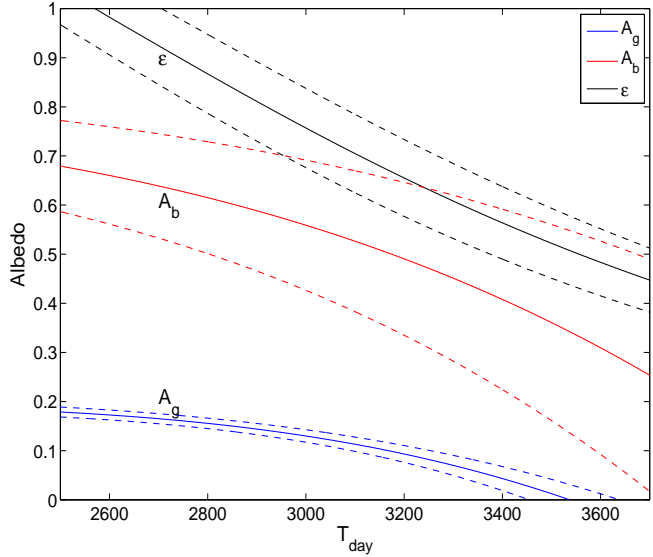}}
\caption{The effective Bond albedo, $A_B$, the geometrical albedo, $A_g$, and $\epsilon$, the energy fraction transferred to the night side, as a function of the day-side temperature. One-sigma errors  are plotted around each curve.
}
\label{AgAB}
\end{figure*}

To proceed, we use the formalism of \citet{cowan11}. They approximate the day-side effective temperature as 

\begin{equation}
T_{\rm D}
=T_{{\rm eff},*}\left(\frac{R_*}{a}\right)^{1/2}
\left(1-A_B\right)^{1/4}
\left(\frac{2}{3}-\frac{5}{12}\epsilon\right)^{1/4} .
\end{equation}
and that of the night side as
\begin{equation}
T_{\rm N}
=T_{{\rm eff},*}\left(\frac{R_*}{a}\right)^{1/2}
\left(1-A_B\right)^{1/4}
\left(\frac{\epsilon}{4}\right)^{1/4} .
\end{equation}
In these expressions $T_{{\rm eff},*}$ is the stellar effective temperature. Two parameters describe the energy budget of the planetary day side ---  $A_B$ is the "effective" Bond albedo, which represents the fraction of the total energy of the star reaching the planetary surface that is not re-emitted thermally by the planetary atmosphere, and $\epsilon$ is the fraction of the energy absorbed by the day side and re-emitted thermally that is transferred to the night side. The estimation of the night-side temperature gives us a strong constraint on the day-side temperature and the fraction of the energy transferred to the night side. This is done by writing 
 \begin{equation}
T_{\rm D}
=T_{\rm N}
\left(\frac{8}{3\epsilon}-\frac{5}{3}\right)^{1/4} \ ,
\end{equation}
and using the derived $T_{\rm N}$ to calculate the dependence of $T_{\rm D}$ on $\epsilon$. We can then derive the Bond albedo for a given $T_{\rm D}$. This is depicted in Figure~\ref{AgAB}. We also plotted in this figure the geometrical albedo, $A_g$, as a function of $T_{\rm D}$, as was done, for example, by \citet{demory11} and \citet{santerne11}, and the corresponding $\epsilon$ value.  

%---------------------------
% Figure 7 %
%---------------------------

The figure suggests that if indeed the derived $T_{\rm N}$ is correct, $A_B$ is substantially larger than $A_g$ in the Kepler band  \citep[e.g.,][]{cowan11} for most cases.

%===============================
%section 5
\section{Discussion}
%===============================

The analysis we present here, based on the beaming modulation detected in the Q2 and Q3 publicly available data, suggests that KOI-13.01 is a massive planet, with a mass in the range of 8--12 $M_{Jup}$. 
The mass derived from the ellipsoidal modulation is not very useful, as the expected amplitude is derived from first-order approximations \citep[e.g.,][]{morris85,morris93}, and therefore the degree of its applicability to hot rapidly rotating stars like KOI-13A \citep{szabo11} is not clear.
When more stars with beaming and ellipsoidal modulations are detected \citep[][]{carter11} we expect the theory to be better tuned and become more accurate. 

\begin{figure*}
\centering
\resizebox{12cm}{10cm}
{\includegraphics{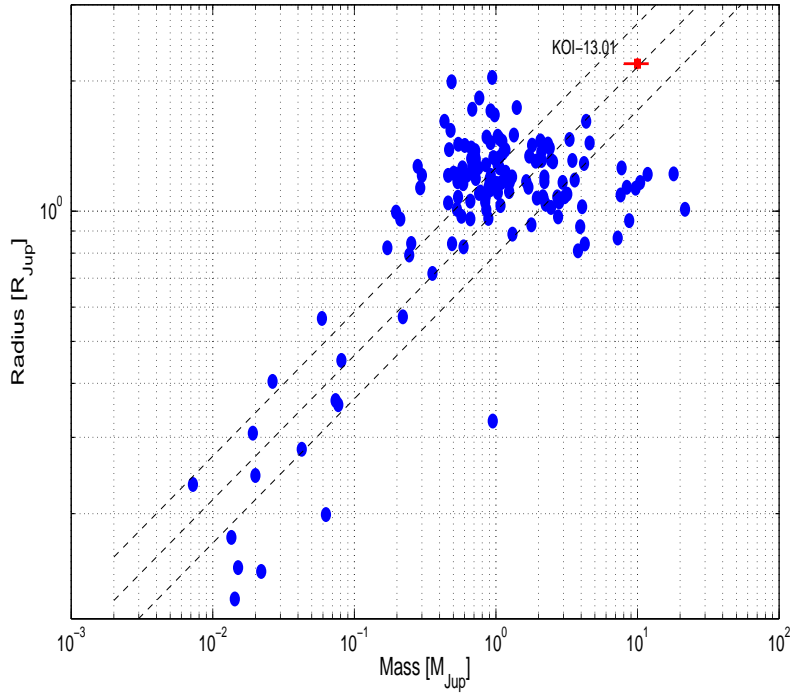}}
\caption{The radii of the transiting planets as a function of their masses, taken from http://www.exoplanet.eu. KOI-13.01 is plotted as a red square. The dashed lines are constant density lines, of 0.5, 1 and 2 Jupiter density.}
\label{RvsM}
\end{figure*}

As of July 2011, KOI-13.01 has the largest radius known for an exoplanet. 
To locate its position in the radius-mass parameter plane we plot the known transiting planets in Figure~\ref{RvsM}, showing that  KOI-13.01 has indeed the largest known radius. It seems as if the upper envelope of the planetary radius reaches its maximum slightly below Jupiter mass.   
If our estimate for its mass is correct, KOI-13.01 is substantially larger than planets with similar masses, but its density is similar to that of Jupiter itself. 
Much theoretical work has been done to explain the large radii 
of some planets \citep[e.g.,][]{guillot02,baraffe04,chabrier07,fortney07,miller09}.
It would be interesting to find observational indications whether KOI-13.01's large radius is associated with its large if not largest insolation \citep[e.g.,][]{enoch11} and/or the relatively large mass of its host star and/or the young age of the system, estimated by \citet{szabo11} to be 0.7 Gyrs. 
  
%---------------------------
% Figure 8 %
%---------------------------

We have discovered additional periodicities in the light curve of KOI-13, with amplitudes of 30 ppm or less and frequencies in the range of 0.5--2 d$^{-1}$. We suspect that the origin of these modulations is stellar. 

The analysis suggests that the secondary occultation depth might be deeper than the reflection peak-to-peak modulation. If true, this is a way to estimate the thermal emission of the  planetary night side of KOI-13.01 \citep[e.g.,][]{cowan11}. We estimate its effective temperature to be $2600\pm150$ K, assuming the detected difference is real, and using a simplistic black-body emissivity approximation. 
We derive the planetary geometrical and effective Bond albedo, the latter representing the fraction of the total energy of the star reaching the planetary surface that is not re-emitted thermally by the planetary atmosphere, as a function of the day-side temperature. Our analysis suggests that the Bond albedo is probably substantially larger than the geometrical albedo. This might indicate that some of the stellar photons not seen in the Kepler band do not share the same processes as the optical photons. For example, UV photons, beyond the Balmer limit, might be absorbed by Hydrogen atoms at the first excited state in the upper layers of the atmosphere and their energy might not be directed to heat the planetary atmosphere. 

Unfortunately, the detection of the night-side thermal emission was not highly significant. With Kepler additional data, beyond Q2 and Q3,  one should be able to determine with higher significance and precision the magnitude of the beaming, ellipsoidal and reflection effects, and compare the secondary depth with the phase modulation amplitude. 

KOI-13 is one of the very few A-type stars known to host a hot Jupiter \citep{desort10}. More cases like KOI-13 might present a unique opportunity to shed some light on the population of giant planets around stars more massive than most of the known planet-hosting stars, and compare them to the planets around G- and K-type stars. Right now this is done mainly by RV search of a sample of sub-giants that are believed to be evolved A-type stars \citep[e.g.,][]{bowler10}. Those sub-giants might not be able to retain hot Jupiters with short periods like that of KOI-13.01. We assume that many non-eclipsing KOI-13.01-type planets are lurking in the Kepler and CoRoT data. In fact we are searching in both data sets for such systems with the BEER algorithm \citep{faigler11a,faigler11b}.

The estimated mass of KOI-13.01 might justify a recognition of its planetary nature, even without confirmation by follow-up RV observations. This was done already, for example, for Kepler-9d \citep{torres11}, Kepler-10c \citep{fressin11}, and Kepler-11g \citep{lissauer2011}. It seems as if the recent large number of transit detections, especially of low-mass planets and early-type host stars, pushes us towards a new terminology of confirmed planets.   

%============================
% Acknowledgements 
%============================

\begin{acknowledgements}
We thank Avi Shporer for helpful discussion and the referee for illuminating comments.\\
We feel deeply indebted  to the team of the Kepler spacecraft, who built and is maintaining this mission, enabling us to search and analyze their unprecedented accurate photometric data.\\
All the photometric data presented in this paper were obtained from the 
Multimission Archive at the Space Telescope Science Institute (MAST). 
STScI is operated by the Association of Universities for Research in 
Astronomy, Inc., under NASA contract NAS5-26555. Support for MAST for 
non-HST data is provided by the NASA Office of Space Science via grant 
NNX09AF08G and by other grants and contracts.\\
This research was supported by the ISRAEL SCIENCE FOUNDATION (grant No.
655/07).
\end{acknowledgements}

%_____________________________bibliography________________________

{}

\begin{thebibliography}{}

\bibitem[Ahmed et al.(1974)]{ahmed74}
Ahmed, N., T. Natarajan, T., \& K. R. Rao, K.R.\ 1974, IEEE Trans. Computers, 90

\bibitem[Anderson et al.(2011)]{anderson11} 
Anderson, D.~R., et al.\ 2011, arXiv:1101.5620

\bibitem[Baraffe et al.(2004)]{baraffe04} 
Baraffe, I., Selsis, F., Chabrier, G., Barman, T.~S., Allard, F., Hauschildt, P.~H., \& Lammer, H.\ 2004, \aap, 419, L13 

\bibitem[Bowler et al.(2010)]{bowler10} 
Bowler, B.~P., et al.\ 2010, \apj, 709, 396

\bibitem[Brown et al.(2011)]{brown11} 
Brown, T.~M., Latham, D.~W., Everett, M.~E., \& Esquerdo, G.~A.\ 2011, arXiv:1102.0342

\bibitem[Borucki et al.(2010)]{borucki10} 
Borucki, W.~J., et al.\ 2010, Science, 327, 977

\bibitem[Borucki et al.(2011)]{borucki11} 
Borucki, W.~J., et al.\ 2011, \apj, 736, 19 

\bibitem[Carter et al.(2011)]{carter11} 
Carter, J.~A., Rappaport, S., \& Fabrycky, D.\ 2011, \apj, 728, 139

\bibitem[Castelli \& Kurucz(2004)]{castelli04} 
Castelli, F., \& Kurucz, R.~L.\ 2004, arXiv:astro-ph/0405087 

\bibitem[Chabrier \& Baraffe(2007)]{chabrier07} 
Chabrier, G., \& Baraffe, I.\ 2007, \apjl, 661, L81

\bibitem[Cowan \& Agol(2011)]{cowan11} 
Cowan, N.~B., \& Agol, E.\ 2011, \apj, 729, 54

\bibitem[Deleuil et al.(2008)]{deleuil08}
Deleuil, M., et al.\ 2008, \aap, 491, 889

\bibitem[Demory et al.(2011)]{demory11} 
Demory, B.-O., et al.\ 2011, \apjl, 735, L12

\bibitem[Desort et al.(2010)]{desort10} 
Desort, M., Lagrange, 
A.-M., Galland, F., Udry, S., \& Mayor, M.\ 2010, EAS Publications Series, 41, 99

\bibitem[Enoch et al.(2011)]{enoch11} 
Enoch, B., et al.\ 2011, \mnras, 410, 1631

\bibitem[Etzel(1980)]{etzel80} 
Etzel, P. B. 1980, {\small EBOP} User's Guide, 3rd edn., UCLA Astronomy and Astrophysics

\bibitem[Faigler \& Mazeh(2011)]{faigler11a} 
Faigler, S., \& Mazeh, T.\ 2011, \mnras, 1106 

\bibitem[Faigler, Mazeh et al.(2011)]{faigler11b} 
Faigler, S., Mazeh, T. et al.\ 2011, accepted for publication \apj, arXiv:1110.2133 

\bibitem[Fortney et al.(2007)]{fortney07} 
Fortney, J.~J., Marley, M.~S., \& Barnes, J.~W.\ 2007, \apj, 659, 1661

\bibitem[Fortney et al.(2008)]{fortney08} 
Fortney, J.~J., Lodders, K., Marley, M.~S., \& Freedman, R.~S.\ 2008, \apj, 678, 1419

\bibitem[Fressin et al.(2011)]{fressin11} 
Fressin, F., et al.\ 2011, arXiv:1105.4647

\bibitem[Galland et al.(2005)]{galland05}
Galland, F., Lagrange, A.-M., Udry, S., Chelli, A., Pepe, F., Queloz, D., Beuzit, J.-L., \& Mayor, M.\ 2005, \aap, 443, 337 

\bibitem[Guillot \& Showman(2002)]{guillot02} 
Guillot, T., \& Showman, A.~P.\ 2002, \aap, 385, 156 

\bibitem[Hartman et al.(2011)]{hartman11} 
Hartman, J.~D., et al.\ 2011, arXiv:1106.1212

\bibitem[Lissauer et al.(2011)]{lissauer2011} 
Lissauer, J.~J., et al.\ 2011, \nat, 470, 53

\bibitem[Loeb \& Gaudi(2003)]{loeb03} 
Loeb, A., \& Gaudi, B.~S.\ 2003,\apjl, 588, L117

\bibitem[Mandel \& Agol(2002)]{mandel02} 
Mandel, K., \& Agol, E.\ 2002, \apjl, 580, L171

\bibitem[Mazeh(2008)]{mazeh08} Mazeh, T.\ 2008, EAS 
Publications Series, 29, 1 

\bibitem[Mazeh \& Faigler(2010)]{mazeh10} 
Mazeh, T., \& Faigler, S.\ 2010, \aap, 521, L59

\bibitem[Miller et al.(2009)]{miller09} 
Miller, N., Fortney, J.~J., \& Jackson, B.\ 2009, \apj, 702, 1413 

\bibitem[Morris(1985)]{morris85} 
Morris, S.~L.\ 1985, \apj, 295, 143

\bibitem[Morris \& Naftilan(1993)]{morris93} 
Morris, S.~L., \& Naftilan, S.~A.\ 1993, \apj, 419, 344 

\bibitem[P{\'a}l et al.(2008)]{pal08}
 P{\'a}l, A., et al.\ 2008, \apj, 680, 1450 

\bibitem[\protect\astroncite{{Popper} and {Etzel}}{1981}]{popper81}
{Popper}, D.~M. and {Etzel}, P.~B. 1981, \aj, 86, 102

\bibitem[Rowe et al.(2011)]{rowe11} 
Rowe, J., Borucki, W.~J., 
Howell, S.~B., Gilliland, R.~L., Buchhave, L.~A., Batalha, N.~M., Latham, 
D.~W., \& Kepler Science Team 2011, Bulletin of the American Astronomical Society, 43, \#103.04

\bibitem[Rybicki \& Lightman(1979)]{rybicki79}
Rybicki, G. B., \& Lightman, A. P.\ 1979, Radiative Processes in Astrophysics (New York: Wiley)

\bibitem[Santerne et al.(2011)]{santerne11} 
Santerne, A., Bonomo, A.~S., H{\'e}brard, G., Deleuil, M., Moutou, C., Almenara, J.~-., Bouchy, 
F., \& D{\'{\i}}az, R.~F.\ 2011, arXiv:1108.0550

\bibitem[Scargle(1982)]{scargle82} 
Scargle, J.~D.\ 1982, \apj, 263, 835 

\bibitem[Snellen et al.(2009)]{snellen09} 
Snellen, I.~A.~G., de Mooij, E.~J.~W., \& Albrecht, S.\ 2009, \nat, 459, 543

\bibitem[Szab{\'o} et al.(2011)]{szabo11} 
Szab{\'o}, G.~M., et al.\ 2011, \apjl, 736, L4 

\bibitem[Tamuz et al.(2006)]{tamuz06} 
Tamuz, O., Mazeh, T., \& North, P.\ 2006, \mnras, 367, 1521

\bibitem[Torres et al.(2011)]{torres11} 
Torres, G., et al.\ 2011, \apj, 727, 24

\bibitem[van Kerkwijk et al.(2010)]{vankerkwijk10} 
van Kerkwijk, M.~H., Rappaport, S.~A., Breton, R.~P., Justham, S., Podsiadlowski,
P., \& Han, Z.\ 2010, \apj, 715, 51

\bibitem[Welsh et al.(2010)]{welsh10} 
Welsh, W.~F., Orosz, J.~A., Seager, S., Fortney, J.~J., Jenkins, J.,
Rowe, J.~F., Koch, D., \& Borucki, W.~J.\ 2010, \apjl, 713, L145

\bibitem[Wilson \& Devinney(1971)]{wilson71} 
Wilson, R.~E., \& Devinney, E.~J.\ 1971, \apj, 166, 605 

\bibitem[Zucker, Mazeh \& Alexander (2007)]{zucker07} 
Zucker, S., Mazeh, T., \& Alexander, T.\ 2007, \apj, 670, 1326 

\end{thebibliography}
\end{document}